\documentclass[preprint,showpacs,preprintnumbers,amsmath,amssymb]{revtex4}
\usepackage[dvips]{graphicx}
\usepackage{amsmath}
\usepackage{bm}

\begin{document}

\preprint{BA-02-17}

\title{Unifying flipped $SU(5)$ in five dimensions}

\author{S.M. Barr}
\email{smbarr@bxclu.bartol.udel.edu}
\author{Ilja Dorsner}
\email{dorsner@physics.udel.edu}
\affiliation{
Bartol Research Institute\\
University of Delaware\\
Newark, DE 19716}

\begin{abstract}
It is shown that embedding a four-dimensional flipped $SU(5)$ model
in a five-dimensional $SO(10)$ model, preserves the best features of
both flipped $SU(5)$ and $SO(10)$. The missing
partner mechanism, which naturally achieves both doublet-triplet
splitting and suppression of $d=5$ proton decay operators, is
realized as in flipped $SU(5)$, while the gauge couplings are
unified as in $SO(10)$. The masses of down quarks and charged leptons,
which are independent in flipped $SU(5)$, are related by the $SO(10)$.
Distinctive patterns of quark and lepton masses can result.
The gaugino mass $M_1$ is independent of $M_3$ and $M_2$, which are
predicted to be equal.
\end{abstract}

\pacs{12.10.Dm,12.60.Jv}

\maketitle

\section{Introduction}
A beautiful feature of flipped $SU(5)$ \cite{Barr:1981qv,Derendinger:1983aj,
Antoniadis:1987dx} is that it provides a natural setting for the missing partner mechanism.
This mechanism, when implemented in flipped
$SU(5)$, not only solves the doublet-triplet splitting problem but
also allows one to avoid entirely the Higgsino-mediated proton decay that
is such a difficulty for supersymmetric grand unified theories (SUSY GUTs). On the
other hand, flipped $SU(5)$ gives up one of the most attractive features of
grand unification, namely unification of gauge couplings, because it is
based on the group $SU(5) \times U(1)$. Another drawback of flipped $SU(5)$
models is that the masses of down quarks and charged leptons come from
different operators, so that one
does not obtain the relation $m_b(M_{GUT}) = m_{\tau}(M_{GUT})$.
The unification of gauge couplings and relations between down quark masses
and charged lepton masses could be recovered by embedding the group
$SU(5) \times U(1)$ in the simple group $SO(10)$. However, in that case,
the missing partner mechanism no longer works, since the partner that was
missing from the $SU(5)$ multiplets is present in the larger $SO(10)$
multiplets.

One thus has somewhat of a quandary. The point of this paper is that
a way out of this quandary is provided by unification in five dimensions.
We show that if the group $SO(10)$ in five dimensions is broken
by orbifold compactification to the group $SU(5) \times U(1)$
in four dimensions it is possible to have at the same time the good
features of both flipped $SU(5)$ and of $SO(10)$. The essential reason is
that if $SO(10)$ is broken by the orbifold compactification then
the fields of the effective four-dimensional theory need not
be in complete $SO(10)$ multiplets. This means that at the four-dimensional
level the famous missing partners can still be missing and the doublet-triplet
splitting can be achieved without the dangerous Higgsino-mediated
proton decay. On the other hand, because there is $SO(10)$ at the
five-dimensional level, there is approximate unification of gauge couplings,
and there is also the possibility of getting $SO(10)$-like Yukawa
couplings for the quarks and leptons.

By now there are many models that use orbifold compactification
of extra dimensions to break grand unified symmetries.
The first such models \cite{Kawamura:1999nj,Kawamura:2000ev,Kawamura:2000ir,
Altarelli:2001qj,Kobakhidze:2001yk,Hall:2001pg,Hebecker:2001wq}
showed that with one extra dimension
it is possible to construct $SU(5)$ models which have natural
doublet-triplet splitting and no problem with the $d=5$ proton decay
operators that plague four-dimensional SUSY GUTs.
The breaking of grand unified symmetries by orbifold
compactification of a single extra dimension does not reduce the
rank of the group \cite{Hebecker:2001jb}. Thus to break $SO(10)$ all
the way to the Standard Model by orbifold compactification requires
at least two extra dimensions. Interesting six-dimensional $SO(10)$
models have been constructed in several papers \cite{Asaka:2001eh,Hall:2001xr,Haba:2002ek}.
However, it is also possible that the breaking from the grand unified
group to the Standard Model is achieved by a combination of orbifold
compactification and the conventional four-dimensional Higgs
mechanism. That allows the construction of realistic $SO(10)$ models
with only a single extra dimension, as was shown by Derm\' \i \v sek and Mafi
\cite{Dermisek:2001hp}. In their model
the theory in the five-dimensional bulk has $\mathcal{N}=1$ supersymmetry
and gauge group $SO(10)$.  Orbifolding breaks $SO(10)$
to the Pati-Salam \cite{Pati:1974yy} symmetry
$SU(4)_c \times SU(2)_L \times SU(2)_R$. The orbifold has two
inequivalent fixed points $O$ and $O'$. On $O$ there is a full
$SO(10)$ symmetry, but on $O'$ only the Pati-Salam group.
On the brane at $O$ the conventional Higgs mechanism breaks $SO(10)$
down to $SU(5)$. Thus the unbroken symmetry in the low-energy theory
in four dimensions is the intersection of $SU(5)$ and the Pati-Salam group,
i.e. the Standard Model group.

The model we shall present is similar in
some ways to that of Derm\' \i \v sek and Mafi but differs from it in
several important respects. Whereas they use orbifold compactification to
break to the Pati-Salam group and Higgs fields on the brane $O$ to break to
$SU(5)$, we shall use orbifold compactification to break to
$SU(5) \times U(1)$ and Higgs fields in the bulk to break the rest of the way
to the Standard Model. They use orbifold breaking to split the doublets
from the triplets, whereas we use the four-dimensional flipped-$SU(5)$
missing partner
mechanism.

\section{Missing partners in four dimensions}
Before we consider higher dimensional theories we shall briefly review
the missing partner mechanism in four-dimensional theories,
showing why it works in flipped $SU(5)$ but not in $SO(10)$.

\subsection{Flipped SU(5)}
First recall what happens in ordinary (i.e. Georgi-Glashow) $SU(5)$ \cite{Georgi:sy}. In ordinary $SU(5)$
the two Higgs doublets of the MSSM, which we shall denote
${\bf 2}$ and $\overline{{\bf 2}}$, have color-triplet
partners, which we shall denote ${\bf 3}$ and $\overline{{\bf 3}}$.
(We use this shorthand notation for Standard Model
representations: ${\bf 2} \equiv (1, 2, -\frac{1}{2})$, $\overline{{\bf 2}}
\equiv (1,2, \frac{1}{2})$, ${\bf 3} \equiv (3,1, \frac{1}{3})$,
$\overline{{\bf 3}} \equiv (\overline{3}, 1, -\frac{1}{3})$.)
These are contained in fundamental and anti-fundamental multiplets
of $SU(5)$: ${\bf 5} = {\bf 2} + {\bf 3}$ and
$\overline{{\bf 5}} = \overline{{\bf 2}} + \overline{{\bf 3}}$.
A combination of an $SU(5)$-singlet mass term and a Yukawa coupling to a Higgs
in the adjoint representation, can (with suitable fine-tuning) give
GUT-scale mass to the triplet partners while leaving the MSSM Higgs
doublets light. This can be represented schematically as

\begin{figure}
\begin{center}
\includegraphics[width=3in]{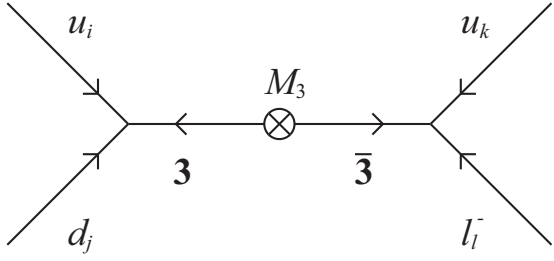}
\end{center}
\caption{\label{dimension5} The kind of graph that gives rise
to $d=5$ proton decay operators.}
\end{figure}

\vspace{0.5cm}
\setlength{\unitlength}{1in}
\begin{picture}(6,1.4)
\thicklines
\put(2.8,1.165){\line(1,0){0.42}}
\parbox[b][1.4in][c]{6in}{
\begin{equation*}\begin{array}{cc}
\left( \begin{array}{c} {\bf 3}\strut\\ {\bf 2}\end{array} \right) & \left(
\begin{array}{c} \overline{{\bf 3}}\strut\\ \overline{{\bf 2}}\end{array} \right) \\
\parallel & \parallel \\
{\bf 5} & \overline{{\bf 5}} \end{array}
\end{equation*}}
\end{picture}

\noindent
where the solid horizontal line represents a large Dirac mass $M_3$
connecting the colored Higgsinos ${\bf 3}$ and $\overline{{\bf 3}}$.
It is well-known that the exchange of these colored Higgsinos gives a
dangerous $d=5$ proton-decay operator, as shown in Fig.~\ref{dimension5}. From the
figure one sees that this proton decay amplitude is proportional to
the mass connecting ${\bf 3}$ to $\overline{{\bf 3}}$. Suppressing
this proton decay therefore requires severing this connection. This can be
done by introducing an extra pair of Higgs multiplets ${\bf 5}'
+ \overline{{\bf 5}}'$, so that the triplets in the unprimed multiplets
get mass not with each other but with the triplets in the
primed multiplets as shown in the following diagram

\vspace{0.5cm}
\setlength{\unitlength}{1in}
\begin{picture}(6,1.4)
\thicklines
\put(2.20,1.165){\line(1,0){0.41}}
\put(3.35,1.165){\line(1,0){0.41}}
\multiput(2.76,0.89)(0.09,0){5}{\line(1,0){0.04}}
\multiput(2.76,1.165)(0.09,0){5}{\line(1,0){0.04}}
\parbox[b][1.4in][c]{6in}{
\begin{equation*}\begin{array}{cccc}
\left( \begin{array}{c} {\bf 3}\strut\\ {\bf 2}\end{array} \right) & \left(
\begin{array}{c} \overline{{\bf 3}}'\strut\\ \overline{{\bf 2}}'\end{array} \right) &
\left( \begin{array}{c} {\bf 3}'\strut\\ {\bf 2}'\end{array} \right) &
\left( \begin{array}{c} \overline{{\bf 3}}\strut\\ \overline{{\bf 2}}\end{array}
\right)\strut\\
\parallel & \parallel & \parallel & \parallel\strut\\
{\bf 5} & \overline{{\bf 5}}' & {\bf 5}' & \overline{{\bf 5}}\end{array}
\end{equation*}}
\end{picture}

\noindent
If the MSSM Higgs doublets are the
unprimed ones, then one sees that their colored partners are not
connected to each other by a mass term, so that the $d=5$ proton-decay
amplitude vanishes. Unfortunately, however, there
is an extra pair of doublets that remains light, namely the primed
ones. The effect of these on the renormalization group
equations would destroy gauge coupling unification.
To give the needed superheavy mass to these doublets one could introduce
a term $M \overline{{\bf 5}}' {\bf 5}'$; however, this would give mass terms
connecting not only ${\bf 2}'$ to ${\bf 2}$ but also
${\bf 3}'$ to $\overline{{\bf 3}}'$ (indicated by dotted lines in the
previous diagram) and thus
indirectly (after the primed triplets were integrated out) reconnecting
${\bf 3}$ to $\overline{{\bf 3}}$ and bringing back the dangerous
$d=5$ proton decay amplitude.

Now let us turn to flipped $SU(5)$ and see how it avoids these problems
very elegantly \cite{Antoniadis:1987dx}.
In flipped $SU(5)$ models one has Higgs fields in
the following representations of $SU(5) \times U(1)$:
$h = {\bf 5}^{-2}$, $\overline{h} = \overline{{\bf 5}}^2$,
$H = {\bf 10}^1$, and $\overline{H} = \overline{{\bf 10}}^{-1}$.
Under the Standard Model group these decompose as follows,
$h = \overline{{\bf 2}} + {\bf 3}$,
$\overline{h} = {\bf 2} + \overline{{\bf 3}}$,
$H = \overline{{\bf 3}} + (3,2, \frac{1}{6}) + (1,1, 0)$,
and $\overline{H} = {\bf 3} + (\overline{3}, 2, -\frac{1}{6})
+ (1,1, 0)$. The Higgs superpotential contains the
terms $h \; H \; H + \overline{h} \; \overline{H} \; \overline{H}$.
When the Standard Model singlets $(1,1, 0)$ in
the $H$ and $\overline{H}$ acquire vacuum expectation
values (VEVs) they break $SU(5) \times U(1)$ down to the Standard Model group
and they also give mass to the triplet Higgs. Schematically,

\vspace{0.5cm}
\setlength{\unitlength}{1in}
\begin{picture}(6,1.4)
\thicklines
\put(1.95,1.165){\line(1,0){0.55}}
\put(3.47,1.165){\line(1,0){0.55}}
\multiput(3,0.7)(0,0.2){4}{\line(0,1){0.1}}
\parbox[b][1.4in][c]{6in}{
\begin{equation*}\begin{array}{cccc}
\left( \begin{array}{c} {\bf 3}\strut\\ \overline{{\bf 2}}\end{array} \right) & \left(
\begin{array}{c} \overline{{\bf 3}}\strut\\ {\rm other}\end{array} \right) &
\left( \begin{array}{c} {\bf 3}\strut\\ \overline{{\rm other}}\end{array} \right) &
\left( \begin{array}{c} \overline{{\bf 3}}\strut\\ {\bf 2}\end{array}
\right)\strut\\
\parallel & \parallel & \parallel & \parallel\strut\\
h & H & \overline{H} & \overline{h}\end{array}
\end{equation*}}
\end{picture}

\noindent
where, for simplicity, $(3,2, \frac{1}{6}) + (1,1, 0) \equiv {\rm other}$.
The triplets in $h$ and $\overline{h}$ get mass with those in
$H$ and $\overline{H}$. However the doublets in $h$ and $\overline{h}$ remain
massless because there are no doublets in $H$ and $\overline{H}$ for them
to mate with --- thus the name ``missing partner mechanism".

At first glance one might worry that the same problem arises here as
in the ordinary $SU(5)$ case discussed previously. The multiplets
${\bf 5}'$ and $\overline{{\bf 5}}'$ there played the same role as
the multiplets $H$ and $\overline{H}$ here. And we saw that one could
not give mass to the doublets in ${\bf 5}'$ and $\overline{{\bf 5}}'$
without reintroducing the dangerous proton decay amplitude. This leads
to the question whether there is not an analogous difficulty in giving mass
to some of the components of $H$ and $\overline{H}$, and specifically to the
$(3,2, \frac{1}{6}) + (1,1,0) + ( \overline{3}, 2, -\frac{1}{6}) +
(1,1,0)$, since here also an explicit mass term $M \overline{H} H$
would reintroduce the problem of proton decay. The beautiful
answer is that these ``other" components of $H$ and $\overline{H}$ do
not have to get mass. Indeed, they {\it must not} get mass, because they
are the goldstone modes that get eaten when $SU(5) \times U(1)$ breaks
to $SU(3) \times SU(2) \times U(1)$. In other words, the fact that
$SU(5) \times U(1)$ breaks to the Standard Model group guarantees that
there is no mass connecting $H$ and $\overline{H}$ and therefore
guarantees the absence of the $d=5$ proton decay amplitude.

\subsection{SO(10)}
Now let us see why embedding flipped $SU(5)$ in $SO(10)$ in four
dimensions destroys the beautiful missing partner solution to
the doublet-triplet splitting and proton decay problems that we have just
reviewed.

In $SO(10)$ the simplest possibility is that the terms
$h \; H \; H + \overline{h} \; \overline{H} \; \overline{H}$ come from the
terms ${\bf 10} \;{\bf 16} \;{\bf 16} + {\bf 10} \; \overline{{\bf 16}}
\; \overline{{\bf 16}}$, where
${\bf 10} = \overline{h} + h$, ${\bf 16} = H + \overline{h}' + {\bf 1}^5$,
and $\overline{{\bf 16}} = \overline{H} + h' + {\bf 1}^{-5}$.
Here $h' = {\bf 5}^3$ and $\overline{h}' = \overline{{\bf 5}}^{-3}$.
The problem is that the doublet partners that were missing from
$H$ and $\overline{H}$ are now present in $\overline{h}'$ and $h'$.

The terms ${\bf 10} \;{\bf 16} \;{\bf 16} + {\bf 10} \;\overline{{\bf 16}}
\;\overline{{\bf 16}}$ contain not only
$h \; H \langle H \rangle + \overline{h} \; \overline{H} \langle
\overline{H} \rangle$ but also $\overline{h} \; \overline{h}' \langle H
\rangle + h \; h' \langle \overline{H} \rangle$. These latter terms mate
the doublet Higgs in $h$ and $\overline{h}$ with those in
$\overline{h}'$ and $h'$, destroying the solution of the doublet-triplet
splitting problem.

A possible remedy to this difficulty suggests itself. One can
have $h$ and $\overline{h}$ come from different ${\bf 10}$s
of $SO(10)$. Let us examine what happens in this case, since it
will be directly relevant to what we shall do in five dimensions
later. Suppose there are two vector Higgs representations, denoted
${\bf 10}_1$ and ${\bf 10}_2$, with couplings
${\bf 10}_1 \; {\bf 16} \; {\bf 16} + {\bf 10}_2 \; \overline{{\bf 16}}
\; \overline{{\bf 16}}$. We write ${\bf 10}_1 = h_1 + \overline{h}_1$
and ${\bf 10}_2 = h_2 + \overline{h}_2$. Suppose that the two light
doublets of the MSSM lie in $h_1$ and $\overline{h}_2$; then the
triplet partners of these light doublets will obtain mass from the terms
$h_1 H \langle H \rangle + \overline{h}_2 \overline{H} \langle
\overline{H} \rangle$. The terms that give superlarge mass to doublets,
and which correspond to those we found troubling before, are
$\overline{h}_1 \overline{h}' \langle H \rangle +
h_2 h' \langle \overline{H} \rangle$. These do {\it not} give superlarge
mass to the MSSM doublets, but to the doublets in $\overline{h}_1$
and $h_2$. Thus, we would appear to have satisfactory doublet-triplet
splitting with no dangerous $d=5$ proton decay, just as in flipped
$SU(5)$.

However, this is not so, for the question arises how the triplets in
$\overline{h}_1$ and $h_2$ are to acquire superheavy mass. It would
seem that the only way is through a mass term connecting them. But
that would have to come from a term $M \overline{h}_1 h_2$, which
in turn comes from $M {\bf 10}_1 {\bf 10}_2$, and this would also
give $M h_1 \overline{h}_2$ and thus superlarge mass to the MSSM doublets.

We see, then, that the missing partner mechanism does not work in
four-dimensional $SO(10)$ theories. However, we shall see that it can work
in five-dimensional $SO(10)$ theories. The crucial difference will be that
orbifold breaking of $SO(10)$ can split the $SO(10)$ Higgs representations.
In particular, in the example we just looked at the troublesome triplets
in $\overline{h}_1$ and $h_2$ can be given Kaluza-Klein masses
by the orbifold compactification while leaving the MSSM doublets
in $h_1$ and $\overline{h}_2$ light.

\section{An $SO(10)$ model in five dimensions}
We now present an $SO(10)$ supersymmetric model in five
dimensions compactified on an $S^1/(Z_2 \times Z'_2)$ orbifold
that yields a realistic supersymmetric flipped $SU(5)$ model in
four dimensions. The breaking of $SU(5) \times U(1)$ down to the
Standard Model gauge group, the doublet-triplet splitting, and
the solution to the problem of $d=5$ proton-decay operators will
all be as in conventional four-dimensional flipped $SU(5)$ models.
Moreover, there will be distinctive flipped $SU(5)$ relationships
among gaugino masses. However, the gauge couplings will be unified
(with some threshold corrections, that can be argued to be small
\cite{Hall:2001pg}) because of the underlying
five-dimensional $SO(10)$ symmetry. And the Yukawa couplings of the
quarks and leptons can have relationships that are similar to what is
found in ordinary $SU(5)$ and $SO(10)$ models rather than in flipped
$SU(5)$.

As already elaborated in Refs.
\cite{Kawamura:1999nj,Kawamura:2000ev,Kawamura:2000ir,
Altarelli:2001qj,Hall:2001pg,Kobakhidze:2001yk,Hebecker:2001wq},
the fifth dimension, being the circle with coordinate $y$ and circumference
$2 \pi R$, is compactified
through the reflection $y \rightarrow -y$ under $Z_2$ and
$y' \rightarrow -y'$ under $Z'_2$ where $y'=y+\pi R/2$. This
identification procedure leaves two fixed points $O$
and $O'$ of $Z_2$ and $Z'_2$ respectively and reduces the physical region
to the interval $y \in [-\pi R/2,0]$. Point $O$ at $y=0$ is the ``visible
brane" while point $O'$ at $y'=0$ is the ``hidden brane".
The compactification scale $1/R \equiv M_C$
is assumed to be close to the scale at which the gauge couplings
unify, i.e. the GUT scale $M_{GUT} \sim 10^{16}$ GeV.

The generic bulk field $\phi(x^\mu,y)$, where $\mu=0,1,2,3$, has definite
parity assignment under $Z_2 \times Z'_2$ symmetry. Taking $P=\pm 1$ to be
parity eigenvalue of the field $\phi(x^\mu,y)$ under $Z_2$ transformation
and $P'=\pm 1$ to be parity eigenvalue under the $Z'_2$ transformation,
a field with $(P,P')=(\pm,\pm)$ can be denoted
$\phi^{PP'}(x^\mu,y)=\phi^{\pm \pm}(x^\mu,y)$. The Fourier series
expansion of the fields $\phi^{\pm \pm}(x^\mu,y)$ yields
\begin{subequations}
\label{Fourier}
\begin{eqnarray}
\phi^{++}(x^\mu,y)&=&\frac{1}{\sqrt{2^{\delta_{n0}}\pi R}}
\sum^{\infty}_{n=0}
\phi^{++(2n)}(x^\mu) \cos \frac{2ny}{R},\\
\phi^{+-}(x^\mu,y)&=&\frac{1}{\sqrt{\pi R}} \sum^{\infty}_{n=0}
\phi^{+-(2n+1)}(x^\mu) \cos \frac{(2n+1)y}{R},\\
\phi^{-+}(x^\mu,y)&=&\frac{1}{\sqrt{\pi R}} \sum^{\infty}_{n=0}
\phi^{-+(2n+1)}(x^\mu) \sin \frac{(2n+1)y}{R},\\
\phi^{--}(x^\mu,y)&=&\frac{1}{\sqrt{\pi R}} \sum^{\infty}_{n=0}
\phi^{--(2n+2)}(x^\mu) \sin \frac{(2n+2)y}{R}.
\end{eqnarray}
\end{subequations}

In the effective theory in four dimensions all the fields in
Eqs.~(\ref{Fourier}) have masses of order $M_C$
except the Kaluza-Klein zero
mode $\phi^{++(0)}$ of $\phi^{++}(x^\mu,y)$, which remains massless.
Moreover, fields $\phi^{-\pm}(x^\mu,y)$ vanish on the visible brane and
fields $\phi^{\pm-}(x^\mu,y)$ vanish on the hidden brane.

In our model, we assume that gauge fields and gauge-non-singlet
Higgs fields exist in the five-dimensional bulk, while the quark
and lepton fields and certain gauge-singlet Higgs fields exist
only on the visible brane at $O$. The gauge fields in the bulk are
of course in a vector supermultiplet of 5d supersymmetry that is an
adjoint representation of $SO(10)$. We will denote it by ${\bf 45}_g$,
where the subscript `$g$' stands for `gauge'.
The gauge-non-singlet Higgs fields in the bulk are in hypermultiplets
of 5d supersymmetry and consist of two tens of $SO(10)$, denoted
${\bf 10}_{1H}$ and ${\bf 10}_{2H}$, and a spinor-antispinor pair of
$SO(10)$ denoted ${\bf 16}_H$ and $\overline{{\bf 16}}_H$.
The subscript `$H$' indicates a Higgs field.

The vector supermultiplet ${\bf 45}_g$ decomposes
into a vector multiplet $V$ and a chiral multiplet $\Sigma$
of $\mathcal{N}= 1$ supersymmetry in four dimensions.
Each hypermultiplet splits into two left-handed chiral
multiplets $\Phi$ and $\Phi^c$, having opposite gauge quantum numbers.
Under the $SU(5) \times U(1)$ subgroup the $SO(10)$ representations
decompose as follows:
${\bf 45} \rightarrow {\bf 24}^0+{\bf 10}^{-4}+\overline{{\bf 10}}^4+
{\bf 1}^0$; ${\bf 10} \rightarrow {\bf 5}^{-2}+\overline{{\bf 5}}^2$;
${\bf 16} \rightarrow {\bf 10}^1 + \overline{{\bf 5}}^{-3} + {\bf 1}^5$;
and $\overline{{\bf 16}} \rightarrow \overline{{\bf 10}}^{-1} +
{\bf 5}^3 + {\bf 1}^{-5}$. With these facts in mind we shall now discuss the
transformation of the various fields under the $Z_2 \times Z'_2$ parity
transformations.

The first $Z_2$ symmetry (the one we denote as unprimed)
is used to break supersymmetry to $\mathcal{N}=1$ in four-dimensions.
($\mathcal{N}=1$ in five dimensions is equivalent to $\mathcal{N}=2$ in
four dimensions; so we are breaking half the supersymmetries.) To do this
we assume that under $Z_2$ the $V$ is even, $\Sigma$ is odd,
$\Phi$ are even, and $\Phi^c$ are odd. The $Z_2'$ is used to break
$SO(10)$ down to $SU(5) \times U(1)$. The ${\bf 24}^0$ and ${\bf 1}^0$ of
$V$ are taken to be even under $Z_2'$, while the ${\bf 10}^{-4}$
and $\overline{{\bf 10}}^4 $ are taken to be odd. In ${\bf 10}_{1H}$
the ${\bf 5}^{-2}$ are taken to be even and the
$\overline{{\bf 5}}^2$ odd, whereas in ${\bf 10}_{2H}$ the
parities are taken to be the reverse, ${\bf 5}^{-2}$ odd and
$\overline{{\bf 5}}^2$ even. All told we have
\begin{subequations}
\label{parity}
\begin{eqnarray}
{\bf 45}_g & = & V^{++}_{{\bf 24}^0} + V^{++}_{{\bf 1}^0} +
V^{+-}_{{\bf 10}^{-4}} + V^{+-}_{\overline{{\bf 10}}^4} + \Sigma^{-+}_{{\bf 24}^0} + \Sigma^{-+}_{{\bf 1}^0} +
\Sigma^{--}_{{\bf 10}^{-4}} + \Sigma^{--}_{\overline{{\bf 10}}^4} \\
{\bf 10}_{1H} & = & \Phi^{++}_{{\bf 5}_1^{-2}} +
\Phi^{+-}_{\overline{{\bf 5}}_1^2} + \Phi^{c --}_{\overline{{\bf 5}}_1^2} + \Phi^{c-+}_{{\bf 5}_1^{-2}} \\
{\bf 10}_{2H} & = & \Phi^{+-}_{{\bf 5}_2^{-2}} +
\Phi^{++}_{\overline{{\bf 5}}_2^2} + \Phi^{c -+}_{\overline{{\bf 5}}_2^2} + \Phi^{c--}_{{\bf 5}_2^{-2}} \\
{\bf 16}_H & = & \Phi^{++}_{{\bf 10}^1} + \Phi^{+-}_{\overline{{\bf 5}}^{-3}}
+ \Phi^{+-}_{{\bf 1}^5} + \Phi^{c--}_{\overline{{\bf 10}}^{-1}} + \Phi^{c-+}_{{\bf 5}^3}
+ \Phi^{c-+}_{{\bf 1}^{-5}} \\
\overline{{\bf 16}}_H & = &
\Phi^{++}_{\overline{{\bf 10}}^{-1}} + \Phi^{+-}_{{\bf 5}^3}
+ \Phi^{+-}_{{\bf 1}^{-5}} + \Phi^{c--}_{{\bf 10}^1} + \Phi^{c-+}_{\overline{{\bf 5}}^{-3}}
+ \Phi^{c-+}_{{\bf 1}^5}
\end{eqnarray}
\end{subequations}
Massless zero modes of the Kaluza-Klein towers
exist only for fields with $Z_2 \times Z_2'$ parity $++$. This
includes $\Phi^{++}_{{\bf 5}^{-2}_1}$, $\Phi^{++}_{\overline{{\bf 5}}^2_2}$,
$\Phi^{++}_{{\bf 10}^1}$, and $\Phi^{++}_{\overline{{\bf 10}}^{-1}}$.
We will call the zero modes of these components $h_1$, $\overline{h}_2$,
$H$, and $\overline{H}$, respectively, using the same notation we
used in the last section. The $h_1$ and $\overline{h}_2$ contain
the two Higgs doublets of the MSSM and their colored partners.

To understand these parity assignments, we observe the
invariance of the action for the bulk fields \cite{Arkani-Hamed:2001tb}
given by
\begin{eqnarray}
\label{action}
\nonumber
S_5&=&\int{{\rm d}^5x}\Bigg\{\frac{1}{4 k g^2}
{\rm Tr}\bigg[\int{{\rm d}^2\theta W^\alpha W_\alpha}+h.c.\bigg]
\qquad\qquad\qquad\\
\nonumber
&+&\frac{1}{k g^2} \int{{\rm d}^4\theta}{\rm Tr}\bigg[\Big((\sqrt{2}\partial_5+\overline{\Sigma})
{\rm e}^{-V}(-\sqrt{2}\partial_5+\Sigma){\rm e}^V+
\partial_5 {\rm e}^{-V}\partial_5 {\rm e}^V\Big)\bigg]\\
\nonumber
&+&\sum_{i=1}^4 \int{{\rm d}^4\theta}\bigg[\Phi_i^c {\rm e}^V \overline{\Phi}_i^c +
\overline{\Phi}_i {\rm e}^{-V}\Phi_i\bigg]\\
&+&\sum_{i=1}^4 \bigg[\int{{\rm d}^2\theta}\Phi_i^c(\partial_5-\frac{1}{\sqrt{2}}
\Sigma)\Phi_i+{\rm h.c.}\bigg]\Bigg\}
\end{eqnarray}
under $y \rightarrow -y$ reflection with the superfields transforming as
\begin{subequations}
\label{fields}
\begin{eqnarray}
V^a(x^\mu,-y) T^a & = & V^a(x^\mu,y) P T^a P\\
\Sigma^a(x^\mu,-y) T^a & = &- \Sigma^a(x^\mu,y) P T^a P\\
\Phi_i(x^\mu,-y)  & = & \pm P \Phi_i(x^\mu,-y)\\
\Phi^c_i(x^\mu,-y)  & = & \mp P^T \Phi^c_i(x^\mu,-y)
\end{eqnarray}
\end{subequations}
where $V=V^a T^a$, and $\Sigma=\Sigma^a T^a$. The $T^a$s are the generators of $SO(10)$
in the appropriate representation with normalization ${\rm Tr}[T^a T^b]=k \delta^{ab}$,
and $P=P^{-1}$ is the parity operator. The replacement $y \rightarrow y'$ and $P \rightarrow P'$ in Eqs.~(\ref{fields})
specifies the transformation of the superfields under
$y' \rightarrow -y'$ reflection. Finally, defining $P$ and $P'$ through their action on the ${\bf 10}$
of $SO(10)$, we associate $P=\sigma_0 \otimes I$ with the $Z_2$ and
$P'=\sigma_2 \otimes I$ with $Z'_2$, where $I$ and $\sigma_0$ are $5 \times 5$ and $ 2 \times 2$
identity matrices and $\sigma_2$ is the usual Pauli matrix.

Having done with the parity assignment for the bulk fields we can turn our
attention towards the brane physics. On the brane at $O$ we put the
three families of quarks and leptons. Since the gauge symmetry on this
brane is $SO(10)$, these are contained in three chiral superfields
that are spinors of $SO(10)$, which we denote ${\bf 16}_i$,
where $i = 1,2,3$ is the family
index. Later for various reasons we shall introduce some gauge-singlet
superfields on the brane at $O$, but let us first discuss the
interactions of the fields introduced up to this point.

The $Z_2$ parity of fields in the ${\bf 16}_i$ must
be positive. The $Z'_2$ parity, determined by the content of
Eqs.~(\ref{parity}), is ${\bf 16} \rightarrow
{\bf 10}^{1 \pm} + \overline{{\bf 5}}^{-3 \mp} +{\bf 1}^{5 \mp}$,
where ${\bf 10}_i=(Q,D,N)_i$, $\overline{{\bf 5}}_i=(U,L)_i$, and
${\bf 1}_i=(E)_i$. The action for the coupling of the matter fields,
residing on the visible brane, with the Higgs fields, coming from the bulk,
is
\begin{eqnarray}
\label{mattera}
\nonumber
S^{matter}_5 & = & \int{{\rm d}^5x} \, \frac{1}{2} \left[\delta(y)+\delta(y-\pi R)\right] \sqrt{2 \pi R}
\, \int{{\rm d}^2\theta} \, A_{ij} {\bf 16}_i {\bf 16}_j  {\bf 10}_{1H} \\
& + & \int{{\rm d}^5x} \, \frac{1}{2} \left[\delta(y)-\delta(y-\pi R)\right] \sqrt{2 \pi R}
\, \int{{\rm d}^2\theta} \, B_{ij} {\bf 16}_i {\bf 16}_j  {\bf 10}_{2H} + {\rm h.c.},
\end{eqnarray}
where $A_{ij}$ and $B_{ij}$ are Yukawa couplings. Integrating over the
fifth dimension $y$
using the Eqs.~(\ref{Fourier}), and keeping only the terms that involve the Yukawa couplings of
the MSSM Higgs doublets and their triplet partners
we obtain the following Lagrangian in four dimensions
\begin{eqnarray}
\label{matterb}
\nonumber
\mathcal{L}^{matter}_4 & = & \sum^{\infty}_{n=0}\int{{\rm d}^2\theta} \sqrt{\frac{2}{2^{\delta_{n0}}}}
\bigg\{ A_{ij}
\Big[ Q_i D_j \overline{d}^{(2n)}_{1H} + L_i E_j \overline{d}^{(2n)}_{1H} +
\frac{1}{2}Q_i Q_j t^{(2n)}_{1H} + U_i E_j t^{(2n)}_{1H} \Big] \\
&+&B_{ij} \Big[ Q_i U_j d^{(2n)}_{2H} + L_i N_j d^{(2n)}_{2H} +
Q_i L_j \overline{t}^{(2n)}_{2H} + U_i E_j \overline{t}^{(2n)}_{2H} \Big] \bigg\} +{\rm h.c.}
\end{eqnarray}
where $\overline{d}_{1H}^{(2n)}$ and $t_{1H}^{(2n)}$ are the doublet and
triplet contained in $\Phi^{++}_{{\bf 5}^{-2}_1}$ (whose zero mode is $h_1$) and
$d_{2H}^{(2n)}$ and $\overline{t}_{2H}^{(2n)}$ are the doublet and
triplet contained in $\Phi^{++}_{\overline{{\bf 5}}^2_2}$
(whose zero mode is $\overline{h}_2$).
All the remaining terms coming from Eq.~(\ref{mattera}) are found by
the replacement $A_{ij} \leftrightarrow B_{ij}$, $(1H) \leftrightarrow (2H)$,
$(2n) \rightarrow (2n+1)$, and $\delta_{n0}
\rightarrow 1$ in Eq.~(\ref{matterb}).

This represents a minimal set of Yukawa terms, and would lead to the
following relations among the quark and lepton mass matrices:
$M_L = M_D \propto A$ and $M_{\nu}^{Dirac} = M_U \propto B$, with
$A$ and $B$ being completely independent symmetric
matrices. This is different from the relations that arise
with a minimal set of Yukawa terms in four-dimensional models based on
$SO(10)$ or flipped $SU(5)$. In four-dimensional flipped $SU(5)$, the minimal
Yukawa terms give $M_{\nu}^{Dirac} = M_U^T$, where these matrices
are not predicted to be symmetric, and no relation for
$M_L$ and $M_D$. In four-dimensional $SO(10)$, the minimal
Yukawa terms give $M_L = M_D \propto M_{\nu}^{Dirac} = M_U$, with
these matrices predicted to be symmetric.

The Higgs fields, though defined in the bulk, will also couple to each
other on the branes. We assume that the Higgs
coupling on the visible brane is of the form
\begin{eqnarray}
\label{higgs}
\nonumber
S^{higgs}_5 & = & \int{{\rm d}^5x} \, \frac{1}{2} \left[\delta(y)+\delta(y-\pi R)\right] \sqrt{2 \pi R}
\, \int{{\rm d}^2\theta} \, {\bf 10}_{1H} {\bf 16}_H {\bf 16}_H\\
& + & \int{{\rm d}^5x} \, \frac{1}{2} \left[\delta(y)-\delta(y-\pi R)\right] \sqrt{2 \pi R}
\, \int{{\rm d}^2\theta} \, {\bf 10}_{2H} {\bf 16}_H {\bf 16}_H
+ {\rm h.c.}.
\end{eqnarray}
There could also be terms of the form ${\bf 10}_{iH} {\bf 10}_{jH}$,
which would directly produce a GUT-scale $\mu$ term and destroy the gauge 
hierarchy. These must be forbidden by a symmetry. This is not a novel 
requirement introduced by the fact that there are extra dimensions. Terms that
would directly produce a GUT-scale $\mu$ term must also be forbidden in 
four-dimensional unified theories.
For example, in four-dimensional $SU(5)$ theories as well as four-dimensional
flipped $SU(5)$ theories, there are Higgs
multiplets in $\overline{{\bf 5}}$ and ${\bf 5}$, and these must be prevented
from obtaining a superheavy mass term together. Similarly, in four-dimensional 
$SO(10)$ theories the light Higgs doublets are typically in a ${\bf 10}$ of 
Higgs, which must be prevented from acquiring a superheavy self-mass term
\cite{Babu:1993we}. 
The same problem arises also in GUTs in higher dimensions. Generally, some
symmetry must be imposed to protect the gauge hierarchy from such dangerous
terms. We shall assume here that there is a $U(1)'$ of the Peccei-Quinn 
type under which the quark and lepton spinors ${\bf 16}_i$ have charge
$+1$, the Higgs fields ${\bf 10}_{1H}$ and ${\bf 10}_{2H}$ have charge
$-2$, and the Higgs fields ${\bf 16}_H$ and $\overline{{\bf 16}}_H$
have charge $+1$. This approach of using a vector-like symmetry to
prevent a large direct $\mu$ term is used in Ref.~\cite{Dermisek:2001hp}.
A drawback of using that method here, as will be seen later, is that
to generate large Majorana mass terms for the neutrinos without
too large a $\mu$ term being generated by higher-dimension operators,
it will be necessary to assume a hierarchy of $10^{-4}$ between the
$U(1)'$ breaking scale and $M_{GUT}$.

Another way of suppressing direct GUT-scale $\mu$ terms is by means of
a continuous $U(1)_R$ symmetry as in Ref.~\cite{Hall:2001pg}.
In that paper it was found that  
$\mu$ and $\mu B$ parameters of the order of the weak scale could be
generated, without any 
fine-tuning, through the Giudice-Masiero mechanism \cite{Giudice:1988yz}.
We do not pursue other approaches such as that here.

The most general effective action of our theory should also
include brane-localized kinetic terms for the modes of the bulk fields
that have non-vanishing wavefunction on the branes.
Since the symmetry that survives on the hidden brane differs
from the symmetry that governs the interactions on the visible
brane and in the bulk, one might worry that the hidden-brane kinetic
terms with the arbitrary coefficients for the gauge fields would spoil
the gauge coupling unification, and that the hidden-brane kinetic terms
for the Higgs fields could affect the mass matrix prediction that stems from
Eq.~(\ref{mattera}).

As it turns out, the gauge kinetic terms on the hidden brane do not
spoil the gauge coupling unification if the volume of the extra dimension
is large enough \cite{Hall:2001pg}. In that case the arbitrary coefficients
of the gauge kinetic terms on the hidden and the visible brane get diluted
and their contribution to the gauge couplings of the Standard Model can
be neglected. The dominant contribution comes from the universal coefficient
that belongs to the gauge kinetic term in the bulk obeying the full symmetry
of the theory.

The hidden brane kinetic terms for the Higgs fields do not affect
the mass relations $M_L = M_D \propto A$ and $M_{\nu}^{Dirac} = M_U \propto B$.
These hidden-brane terms violate $SO(10)$ but respect $SU(5) \times U(1)$, and
so will have the effect of changing the relative normalization of the
$\overline{{\bf 5}}$ and ${\bf 5}$ of Higgs that are inside the same ${\bf 10}$
of $SO(10)$.
However, the ${\bf 5}$ of Higgs and the $\overline{{\bf 5}}$ of Higgs that
contribute to quark and lepton masses in this model come from
different ${\bf 10}$'s of Higgs anyway. The former comes from ${\bf 10}_{1H}$,
while the latter comes from ${\bf 10}_{2H}$. While the matrices $A$ and $B$ will
be differently affected by the hidden-brane kinetic terms, the predictions that
$M_L = M_D \propto A$ and $M_{\nu}^{Dirac} = M_U \propto B$ are not affected 
by that.
The essential point is that these predictions depend only on the $SU(5)$ 
that is
respected by the hidden-brane kinetic terms and not on the full $SO(10)$.

As noted earlier, the only massless modes
of the Higgs fields are $h_1 \subset \Phi^{++}_{{\bf 5}^{-2}_1} \subset
{\bf 10}_{1H}$, $\overline{h}_2 \subset \Phi^{++}_{\overline{{\bf 5}}^2_2}
\subset {\bf 10}_{2H}$, $H \subset \Phi^{++}_{{\bf 10}^1} \subset
{\bf 16}_H$, and $\overline{H} \subset \Phi^{++}_{\overline{{\bf 10}}^{-1}}
\subset \overline{{\bf 16}}_H$. Therefore, after integrating over the fifth
dimension, the terms in Eq.~(\ref{higgs}) yield in the superpotential
of the low-energy effective theory the terms
$h_1 \; H \; H + \overline{h}_2 \; \overline{H} \; \overline{H}$.
These are just the same terms that are present in conventional four-dimensional
flipped $SU(5)$ models to do the doublet-triplet splitting.

We assume that the $H$ and $\overline{H}$ acquire superlarge vacuum
expectation values that break $SU(5) \times U(1)$ down to the Standard
Model group. The tree-level
scalar potential generated by the terms
$h_1  H  H + \overline{h}_2  \overline{H}  \overline{H}$
is flat in this direction. However, as noted in \cite{Antoniadis:1987dx},
this flatness can be lifted by radiative effects after supersymmetry
is broken. It is also possible that additional
terms in the Higgs superpotential on the visible brane can lead
to a tree-level superpotential that produces the required symmetry
breaking, as we shall see later.

Besides breaking the gauge symmetry from $SU(5) \times U(1)$ down
to $SU(3) \times SU(2) \times U(1)$, the vacuum expectation values of
the fields $H \subset {\bf 16}_H$ and $\overline{H} \subset
\overline{{\bf 16}}_H$ allow masses for the right-handed
neutrinos. Such masses come from effective operators of the form
${\bf 16}_i {\bf 16}_j \overline{{\bf 16}}_H \overline{{\bf 16}}_H$.
However, this product of fields has charge $+4$ under the
symmetry $U(1)'$. Consequently, this symmetry must be spontaneously broken.
It must be broken in such a way as to permit sufficiently large
right-handed neutrino masses without at the same time allowing too large
a $\mu$ parameter (which is the coefficient of the term ${\bf 10}_{1H}
{\bf 10}_{2H}$). This can be done in the following way (which we do not
claim to be unique). Suppose that there are fields $S$ and $\overline{S}$
living on the brane at $O$ that are singlets under $SO(10)$
and that have $U(1)'$ charges $+1$ and $-1$ respectively. In the
superpotential on the brane at $O$ there can be terms of the form
$(\overline{S} S - M^2) X$, where $M = \epsilon M_{GUT}$, with
$\epsilon \ll 1$. These terms force $\langle S \rangle =
\langle \overline{S} \rangle = M$. Let us suppose that on the brane at $O$
there are, in addition to the quark and lepton families in ${\bf 16}_i$,
some leptons ${\bf 1}_i$ ($i=1,2,3$) that are $SO(10)$ singlets but have charge
$-1$ under $U(1)'$. Then the following terms in the superpotential at
$O$ are possible: $C_{ij} {\bf 16}_i {\bf 1}_j \overline{{\bf 16}}_H
\overline{S}/M_{*} + F_{ij} {\bf 1}_i {\bf 1}_j S^2/M_{*}$, where
the dimensionless coefficients $C_{ij}$ and $F_{ij}$ are assumed to be
of order one. The mass $M_{*}$ is an ultraviolet cutoff that specifies
the scale at which new physics (eg. other dimensions beyond five, strings)
become important. We take $M_{*}$ to be close to $M_{GUT}$ but, of
course, somewhat larger. These terms give a mass matrix for the neutrinos
that has the form
\begin{equation}
(\nu_i \quad N^c_i \quad {\bf 1}_i) \left( \begin{array}{ccc}
0 & (M_{\nu}^{Dirac})_{ij} & 0 \\ (M_{\nu}^{Dirac})_{ji}
& 0 & C_{ij} \epsilon \overline{M} \\
0 & C_{ji} \epsilon \overline{M} & F_{ij} \epsilon^2 \overline{M}
\end{array} \right)
\left( \begin{array}{c} \nu_j \\ N^c_j \\ {\bf 1}_j \end{array}
\right),
\end{equation}
where $\overline{M} \equiv M_{GUT}^2/M_{*}$.
(Note that we have taken $\langle {\bf 16}_H \rangle = M_{GUT}$.)
It is clear that the six superheavy neutrinos have masses of order
$\epsilon \overline{M}$, whereas the three light (left-handed)
neutrinos have masses of order $(M_{\nu}^{Dirac})^2/\overline{M}$. Taking
the largest neutrino mass $m_3$
to be about $6 \times 10^{-2}$ eV, as suggested
by the atmospheric neutrino data, and its Dirac mass to be $m_c
\cong 174$ GeV, as suggested by the relation $M_{\nu}^{Dirac} = M_U$
(which would hold in a minimal $SO(10)$ model), one has
that $\overline{M} \sim 10^{15}$ GeV. This accords well
with the assumption that $M_{*}$ is slightly larger than the GUT scale
$M_{GUT} \sim 10^{16}$ GeV.

The reason that we have assumed that the parameter $\epsilon \equiv
\langle S \rangle/M_{GUT}$ is much smaller than one is that it
suppresses certain dangerous operators.
For example, $U(1)'$ allows the effective
operator ${\bf 16}_i {\bf 16}_j {\bf 16}_k {\bf 16}_{\ell}
\overline{S}^4/M_{*}^5$. This gives a $d=5$ proton decay
operator with coefficient of order $\epsilon^4 (1/M_{*})$.
Sufficient suppression of proton decay requires that
$\epsilon \sim 10^{-3}$ to $10^{-4}$. Similarly, $U(1)'$ allows
the operator ${\bf 10}_{1H} {\bf 10}_{2H} S^4/M_{*}^3$.
This gives a $\mu$ parameter for the MSSM doublet Higgs fields that
is of order $\epsilon^4 M_{*}$. Requiring that this be no larger
than the weak scale requires that $\epsilon$ be less than about
$3 \times 10^{-4}$. This corresponds to right-handed neutrino masses
of order $3 \times 10^{11}$ GeV. Such intermediate mass scales
for $M_R$ are good from the point of view of leptogenesis \cite{leptogenesis}.

The same singlet Higgs field $S$ can play a role in generating the
vacuum expectation value for the spinor Higgs fields ${\bf 16}_H$ and
$\overline{{\bf 16}}_H$. Such VEVs, as we have already noted, can arise
due to radiative effects after SUSY breaking. But they can also arise
at tree level from a term in the superpotential on the brane at $O$
of the form $(\lambda \overline{{\bf 16}}_H {\bf 16}_H - S^2) Y$,
where $Y$ is a singlet superfield with $U(1)'$ charge of $-2$, and
$\lambda \sim \epsilon^2$. Note that the $F$-terms of the
fields ${\bf 16}_H$ and $\overline{{\bf 16}}_H$ force $\langle Y \rangle
= 0$, meaning
that there is no mass term linking $\overline{{\bf 16}}_H$ to
${\bf 16}_H$ and thus $\overline{H}$ to $H$.
The $U(1)'$ charge assignments allow the higher dimensional
term $\overline{S}^2 \overline{{\bf 16}}_H {\bf 16}_H/M_{*}$.
This will shift the VEV of $Y$, but the $F$-terms of the
fields ${\bf 16}_H$ and $\overline{{\bf 16}}_H$ still enforce the
condition that there is no mass term linking
$\overline{{\bf 16}}_H$ to ${\bf 16}_H$.

Let us now examine the doublet-triplet splitting and proton decay
problems. The terms $h_1 H \langle H \rangle + \overline{h}_2
\overline{H} \langle \overline{H} \rangle$ will couple the
triplets in $h_1$ and $\overline{h}_2$ to those in
$H$ and $\overline{H}$. The doublets in $h_1$ and $\overline{h}_2$
remain light and are the two doublets of the MSSM.
There is no problem with $d=5$ proton decay, because the
triplet partners of the MSSM Higgs doublets are not connected to each other.
The triplets in $h_1$ and $H$ have no mass terms with
the triplets in $\overline{h}_2$ and $\overline{H}$.
Moreover, there are no unwanted light states contained in the Higgs
multiplets ${\bf 10}_{1H}$, ${\bf 10}_{2H}$, ${\bf 16}_H$,
$\overline{{\bf 16}}_H$. In the zero modes ($h_1$, $\overline{h}_2$,
$H$, and $\overline{H}$), the doublets remain light, the triplets become
superheavy by coupling to the VEVs of $H$ and $\overline{H}$, and the
other gauge-non-singlet fields get eaten by the Higgs mechanism when
$SU(5) \times U(1)$ breaks to the Standard Model group. All the non-zero
modes, of course, have superheavy Kaluza-Klein masses.  This is the
crucial difference with four-dimensional theories in which flipped $SU(5)$
is embedded in $SO(10)$. In
four dimensions, as we saw in the last section, the $SO(10)$ Higgs multiplets
${\bf 10}_{1H}$ and ${\bf 10}_{2H}$ when decomposed
under $SU(5) \times U(1)$ contain not only $h_1$ and $\overline{h}_2$ but also
$\overline{h}_1$ and $h_2$; and these multiplets have triplets that
cannot be given mass without destroying the gauge hierarchy. Here,
however, these extra pieces are all made heavy by the orbifold
compactification, since they do not have parity $++$. Thus it is the
fact that the unification of $SU(5) \times U(1)$ into $SO(10)$ occurs
only in higher dimensions that allows the missing partner mechanism
to be implemented.

We have seen that with what may be called the minimal Yukawa
couplings ${\bf 16}_i {\bf 16}_j (A_{ij} {\bf 10}_{1H}
+ B_{ij} {\bf 10}_{2H})$ this model gives distinctive relations among mass
matrices that are different from those that result in four
dimensional models with minimal Yukawa couplings in either $SO(10)$
or flipped $SU(5)$. In particular, $M_L = M_D$, and $M_{\nu}^{Dirac} =
M_U$, with all these matrices being symmetric. This does give
the desired relation $m_b = m_{\tau}$ at the unification scale,
a result of the fact that flipped $SU(5)$ is embedded in $SO(10)$.
However, this minimal set of Yukawa terms is clearly not enough
to give a realistic model of quark and lepton masses.

Recently it has been found that realistic and simple models of
quark and lepton masses can be constructed using so-called
``lopsided" mass matrices \cite{Babu:1995hr,Sato:1997hv,Albright:1998vf,Irges:1998ax}.
The essential feature of such models is
that the mass matrices of the down quarks and charged leptons
are highly asymmetric and that $M_L \sim M_D^T$. Such a relationship
between $M_L$ and $M_D^T$ is typical of models
with an ordinary $SU(5)$, not flipped
$SU(5)$. However, as we shall now see, because the flipped $SU(5)$
is here embedded in $SO(10)$ at the five-dimensional level, it
is possible to obtain such a lopsided structure.

\begin{figure}
\begin{center}
\includegraphics[width=3in]{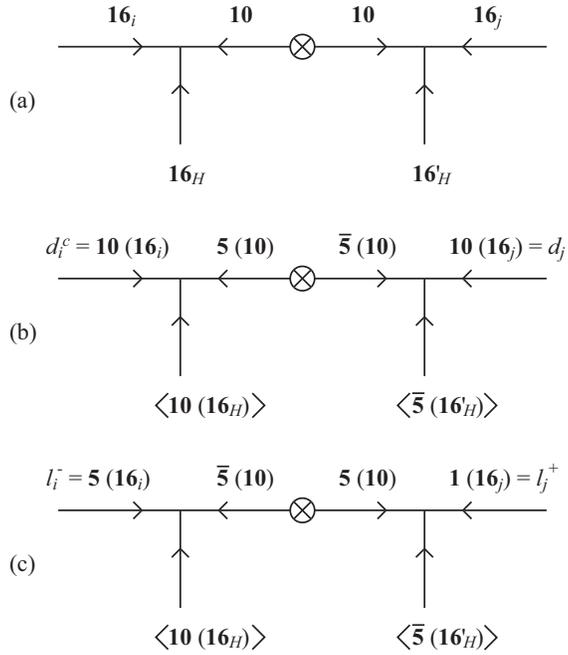}
\end{center}
\caption{\label{masses} (a) A diagram that can give operators producing
``lopsided" contributions to $M_D$ and $M_L$.
(b) A term in its $SU(5) \times U(1)$ decomposition
that contributes to $M_D$.
(c) A term in its $SU(5) \times U(1)$ decomposition
that contributes to $M_L$.}
\end{figure}

Suppose that one introduces on the visible brane not only spinors
of quarks and leptons, but $SO(10)$ vectors as well. And suppose that
there is in the bulk a spinor Higgs field ${\bf 16}'_H$ that has a weak-doublet
component that contributes to the breaking of the electroweak
symmetry. Then a diagram like that shown in Fig.~\ref{masses}(a) may be possible.
When decomposed under the $SU(5) \times U(1)$ subgroup, this diagram
contains the two diagrams shown in Figs.~\ref{masses}(b) and \ref{masses}(c). It is easy to
see that these give contributions to $M_L$ and $M_D$ that are
asymmetric and that are transposes of each other, just as needed
to build ``lopsided" models. The reason for this is that the diagram
in Fig.~\ref{masses}(a) directly depends only on the GUT-scale breaking done
by the ${\bf 16}_H$ and not on that coming from orbifold compactification.
The point is that the ${\bf 16}_H$ VEV by itself would only break
$SO(10)$ down to the Georgi-Glashow $SU(5)$. (It is the orbifold
compactification that breaks $SO(10)$ to the flipped
$SU(5) \times U(1)$ group.) That is why
this diagram leads to the kind of mass contributions that one expects
from ordinary Georgi-Glashow $SU(5)$. This reasoning also shows that
in order to introduce into the mass matrices contributions that break
Georgi-Glashow $SU(5)$ it is necessary that the mass-splittings produced by
the orbifold compactification be involved. For example, by mixing
quarks and leptons that are on the visible brane with quarks and leptons
in the bulk, it should be possible to break the (bad) minimal
$SU(5)$ relations $m_s = m_{\mu}$ and $m_d = m_e$.

\section{Gaugino mediated supersymmetry breaking}
In this section we address the issue of how to break $\mathcal{N}=1$ supersymmetry
of our model below the compactification scale $M_C$. As it turns out, the solution
allows the construction of viable SUSY breaking model that can easily satisfy present
experimental constraints.

It is well known that the models with visible and hidden branes separated by extra
dimension(s) naturally accommodate breaking of supersymmetry via gaugino mediation
\cite{Kaplan:1999ac,Chacko:1999mi}. The basic idea behind gaugino mediation in the
models based on the orbifold compactification is as follows. The source of the SUSY
breaking is localized at the hidden brane. It couples directly to the gauginos at that
brane providing them with nonzero masses. If the gauge symmetry at the hidden brane
is reduced with respect to the bulk gauge symmetry this coupling induces non-universal
gaugino masses. For example, if the bulk symmetry is $SO(10)$ and hidden brane symmetry
is flipped $SU(5)$ one obtains $M_3=M_2\neq M_1$. Here, $M_1$, $M_2$, and $M_3$ represent
gaugino masses of the MSSM.

Following in the footsteps of \cite{Dermisek:2001hp}, we take the source of the SUSY
breaking to be a flipped $SU(5)$ singlet chiral superfield $Z$ localized on the
hidden brane with the VEV
\begin{equation}
\langle Z \rangle=\theta^2 F_Z.
\end{equation}
The gaugino masses originate from the non-renormalizable operators at the hidden brane
of the form
\begin{equation}
\mathcal{L}^{Z}_5=\frac{1}{2}[\delta(y-\pi R/2)+\delta(y+\pi R/2)]
\int{{\rm d}^2\theta}\Big(\lambda_5^\prime\frac{Z}{M^2_*}W^{i\alpha}
W^i_\alpha+\lambda_1^\prime\frac{Z}{M^2_*}W^\alpha W_\alpha+{\rm h.c.}\Big),
\end{equation}
where the first and the second term under the integral represent the $SU(5)$ and $U(1)$
part of the gauge group respectively. Corresponding gaugino masses generated in this way
are
\begin{equation}
\label{gauginos}
M_{SU(5)}=\frac{\lambda_5^\prime F_Z M_c}{M^2_*}, \qquad M_{U(1)}=\frac{\lambda_1^\prime F_Z M_c}{M^2_*},
\end{equation}
which translates into the following MSSM gaugino masses (we normalize the generators of $SO(10)$
demanding that $k=1/2$)
\begin{equation}
\label{gauginos1}
\frac{M_1}{g_1^2}=\frac{1}{25} \frac{M_{SU(5)}}{g_{SU(5)}^2}+\frac{24}{25} \frac{M_{U(1)}}{g_{U(1)}^2},
\qquad M_2=M_{SU(5)},\qquad M_3=M_{SU(5)}.
\end{equation}
Here $g_{SU(5)}$, and $g_{U(1)}$ are gauge coupling constants of the $SU(5)$ and $U(1)$ gauge groups
respectively, while $g_1$ represents the $U(1)_Y$ gauge coupling constant of the Standard Model
(normalized as in GUTs). The
relations of Eq.~(\ref{gauginos1}), which is valid at the compactification scale $M_C$, show that
the gaugino mass $M_1$ can in general be completely different from the mass $M_2=M_3$ due to their
different origins. Namely, the mass $M_1$ is dominated by the $U(1)$ sector of the theory as oppose
to the masses $M_2$ and $M_3$ that have their origin in the $SU(5)$ part of the theory. We will later
see that this feature of non-universality of gaugino masses allows the construction of the
theory of SUSY breaking that leads to the realistic mass spectrum.

At this point we note that the natural scale for $\sqrt{F_Z}$ is the cutoff scale $M_*$. (For the
reasons that have to do with gauge coupling unification we take $(M_C \sim 10^{16}$ GeV$) < (M_{GUT} = 1.2
\times 10^{16}$ GeV$) < (M_* \sim 10 M_C)$ \cite{Dermisek:2001hp}.) This
implies that masses in Eq.~(\ref{gauginos}) are close to the compactification scale $M_C$
if the dimensionless coefficients $\lambda_1^\prime$ and $\lambda_5^\prime$ are taken to be of
order one. To obtain SUSY breaking masses that are in the TeV range we need to decrease the value
of $F_Z$ in a way that does not involve any fine-tuning. To do that we propose to use the shining
mechanism \cite{Arkani-Hamed:1998,Arkani-Hamed:1999pv} which can reduce the natural scale of $F_Z$
by an exponential factor.

The shining mechanism requires the existance of a source $J$ that is localized on the visible
brane and a massive hypermultiplet in the bulk. The hypermultiplet of mass $m$ is taken to be a
gauge singlet and has couplings with both the source and the superfield $Z$. These couplings can
be arranged in a manner that leaves the superfield $Z$ with the nonzero F-term $F_Z \sim J {\rm exp}(-\pi m R/2)$
after the massive hypermultiplet is integrated out \cite{Arkani-Hamed:1999pv}. If the mass $m$ is taken to be
of order $M_*$ the $\sqrt{F_Z}$ will be of order $10^{12}$ GeV which gives desired TeV scale masses for gauginos in
Eq.~(\ref{gauginos}).

The matter fields in our model reside on the visible brane. Thus, due to the spatial separation between
the branes the soft SUSY breaking scalar masses and trilinear couplings are negligible at the compactification
scale. This is good because the number of the soft SUSY breaking parameters one has to consider is reduced
with respect to the usual set.

There are two additional positive features of the gaugino mediated supersymmetry breaking
models with the non-universal gaugino masses. Firstly, the
renormalization group running of scalar
masses and trilinear couplings between $M_C$ and electroweak scale is significantly
affected by gaugino masses but these contributions, being flavor blind, do not cause
any disastrous flavor violating effects. Secondly, non-universality opens up the
possibility for the deviation from the experimentally disfavored prediction of the
models with universal gaugino mass of stau being the lightest supersymmetric particle
(LSP). (The last statement holds for $M_C<M_{GUT}$ which is exactly the case we have.)

The class of models with non-universal gaugino mediated supersymmetry breaking has
been studied in more details by Baer et al. \cite{Baer:2002by}. Their numerical study
of the allowed region of SUSY parameter space shows that viable models with acceptable
mass spectrum and neutral LSP particle can be obtained. The study includes the case of
completely independent $M_3$, $M_2$, and $M_1$, as well as the case where
$M_1$ is a definite linear combination (determined by group theory)
of $M_2$ and $M_3$. (The former case can be seen as a consequence of
orbifold reduction of $SU(5)$ down to the Standard Model group on the hidden brane
as in Ref.~\cite{Hall:2001pg} and
the latter one follows from the reduction of $SO(10)$ down to the Pati-Salam group
as in Ref.~\cite{Dermisek:2001hp}.) We
have an intermediate scenario where $M_1$ is independent of $M_2$ and $M_3$ which are
made equal due to the $SU(5)$ part of the flipped $SU(5)$. (This possibility was
considered in Ref.~\cite{Hall:2001xr} in the context of a six dimensional $SO(10)$ model.)

\begin{figure}
\begin{center}
\includegraphics[width=5in]{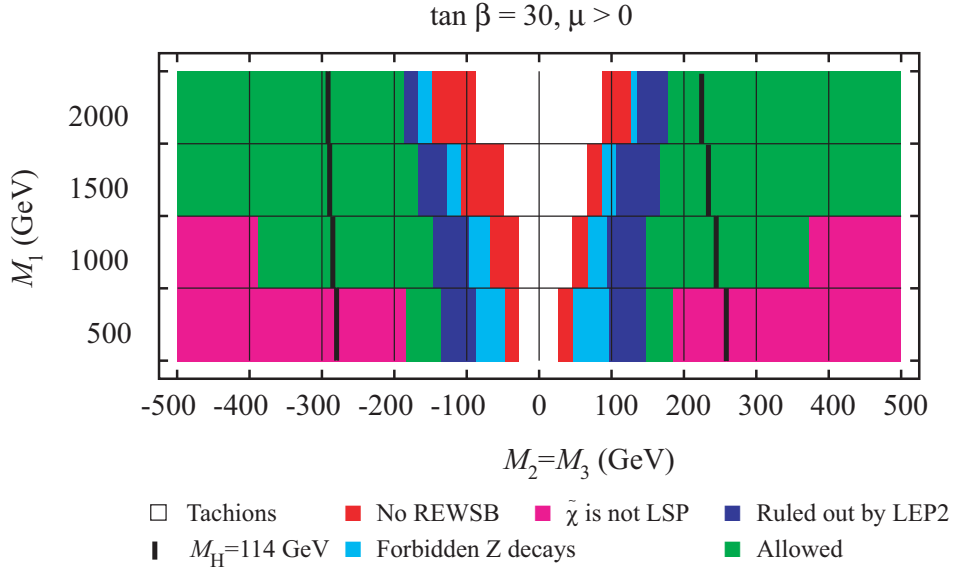}
\end{center}
\caption{\label{region} This diagram represents the results of numerical analysis of
Baer et al. \cite{Baer:2002by} for the case of gaugino mediated SUSY breaking scenario in the
flipped $SU(5)$ setting ($M_2=M_3 \neq M_1$) for $\tan \beta = 30$ and $\mu > 0 $. The allowed
region in $M_1$ vs. $M_2=M_3$ plane is shown in dotted light gray. The excluded regions are white (due to
presence of tachyonic particles in mass spectrum), light gray (due to lack of radiative breakdown of
EW symmetry), gray (due to LEP constraint), dark gray (due to LEP2 constraint), and crossed gray
(due to the fact that charged particle is LSP). Vertical black line is where $M_H=114$ GeV. For a full
discussion of numerical methods and assumptions used in the analysis see Ref.~\cite{Baer:2002by}.}
\end{figure}

It is not difficult to adapt the analysis of Baer et al. to our model to conclude that for
large enough $M_1$ (i.e. $|M_1| >|M_2|,M_2=M_3)$ at the compactification scale $M_C$ a
viable region of parameter space opens up regardless of $\tan \beta $ value yielding realistic
mass spectrum with the LSP being wino-like or a mixture of higgsino and bino. An example of this
behavior is shown in Fig.~\ref{region}.

At the end we observe that if we had the case of $SO(10)$ being reduced on the hidden
brane to the Georgi-Glashow $SU(5)$ with an extra $U(1)$ symmetry we would not only be
prevented from using the simple form of the missing partner mechanism but would also
obtain universal gaugino masses $M_1=M_2=M_3$.

\section{Conclusions}
We have seen that by embedding a four-dimensional flipped $SU(5)$ model
into a five-dimensional $SO(10)$ model the advantages of
flipped $SU(5)$ can be maintained while avoiding its well-known
drawbacks. The two main drawbacks are the loss of unification of gauge
couplings and the loss of the possibility of relating down quark masses
to charged lepton masses, and therefore of obtaining desirable
predictions such as $m_b = m_{\tau}$ and realistic quark and lepton mass
schemes such as those based on ``lopsided" mass matrices.
By embedding $SU(5) \times U(1)$ in $SO(10)$, the unification of gauge
couplings is restored. There are corrections to this unification,
coming for example from gauge kinetic terms on the hidden brane;
however, these have been argued to be small \cite{Hall:2001pg}. The embedding
in $SO(10)$ also yields relationships between the charged lepton and down
quark mass matrices. We have also found that interesting patterns of
quark and lepton masses are possible that are different from
those encountered in four-dimensional grand unified theories, for
example $M_L = M_D \neq M_{\nu}^{Dirac} = M_U$.

Embedding flipped $SU(5)$ in $SO(10)$ in four dimensions is well known
to destroy the missing partner mechanism for doublet-triplet splitting,
which is one of the most elegant features of flipped $SU(5)$.
However, when the unification in $SO(10)$ takes place in higher dimensions
and the breaking to $SU(5) \times U(1)$ is achieved through
orbifold compactification, then the missing partner mechanism can
still operate, as we have shown. One of the advantages of the
missing partner mechanism in flipped $SU(5)$ is that it kills the
dangerous $d=5$ proton decay operators that plague supersymmetric
grand unified theories.

Thus in extra dimensions it is possible to have the best of both worlds,
the best of $SO(10)$ combined with the best of flipped $SU(5)$.
One of the distinctive predictions of the flipped $SU(5)$ scheme
that we have presented is that the gaugino masses will have the
pattern $M_3 = M_2 \neq M_1$. The fact that $M_1$ is independent
of $M_2$ and $M_3$ allows a much larger viable region of parameter
space for the MSSM.

\section{Acknowledgments}
I.~D. thanks R. Derm\' \i \v sek for discussion.

\begin{thebibliography}{99}

\bibitem{Barr:1981qv}
S.~M.~Barr,
Phys.\ Lett.\ B {\bf 112}, 219 (1982).

\bibitem{Derendinger:1983aj}
J.~P.~Derendinger, J.~E.~Kim and D.~V.~Nanopoulos,
Phys.\ Lett.\ B {\bf 139}, 170 (1984).

\bibitem{Antoniadis:1987dx}
I.~Antoniadis, J.~R.~Ellis, J.~S.~Hagelin and D.~V.~Nanopoulos,
Phys.\ Lett.\ B {\bf 194}, 231 (1987).

\bibitem{Kawamura:1999nj}
Y.~Kawamura,
Prog.\ Theor.\ Phys.\  {\bf 103}, 613 (2000)
[arXiv:hep-ph/9902423].

\bibitem{Kawamura:2000ev}
Y.~Kawamura,
Prog.\ Theor.\ Phys.\  {\bf 105}, 999 (2001)
[arXiv:hep-ph/0012125].

\bibitem{Kawamura:2000ir}
Y.~Kawamura,
Prog.\ Theor.\ Phys.\  {\bf 105}, 691 (2001)
[arXiv:hep-ph/0012352].

\bibitem{Altarelli:2001qj}
G.~Altarelli and F.~Feruglio,
Phys.\ Lett.\ B {\bf 511}, 257 (2001)
[arXiv:hep-ph/0102301].

\bibitem{Kobakhidze:2001yk} A.~B.~Kobakhidze,
Phys.\ Lett.\ B {\bf 514}, 131 (2001)
[arXiv:hep-ph/0102323].

\bibitem{Hall:2001pg}
L.~J.~Hall and Y.~Nomura,
Phys.\ Rev.\ D {\bf 64}, 055003 (2001)
[arXiv:hep-ph/0103125].

\bibitem{Hebecker:2001wq}
A.~Hebecker and J.~March-Russell,
Nucl.\ Phys.\ B {\bf 613}, 3 (2001)
[arXiv:hep-ph/0106166].

\bibitem{Hebecker:2001jb}
A.~Hebecker and J.~March-Russell,
Nucl.\ Phys.\ B {\bf 625}, 128 (2002)
[arXiv:hep-ph/0107039].

\bibitem{Asaka:2001eh}
T.~Asaka, W.~Buchmuller and L.~Covi,
Phys.\ Lett.\ B {\bf 523}, 199 (2001)
[arXiv:hep-ph/0108021].

\bibitem{Hall:2001xr}
L.~J.~Hall, Y.~Nomura, T.~Okui and D.~R.~Smith,
Phys.\ Rev.\ D {\bf 65}, 035008 (2002)
[arXiv:hep-ph/0108071].

\bibitem{Haba:2002ek}
N.~Haba, T.~Kondo and Y.~Shimizu,
arXiv:hep-ph/0202191.

\bibitem{Dermisek:2001hp}
R.~Dermisek and A.~Mafi,
Phys.\ Rev.\ D {\bf 65}, 055002 (2002)
[arXiv:hep-ph/0108139].

\bibitem{Pati:1974yy}
J.~C.~Pati and A.~Salam,
Phys.\ Rev.\ D {\bf 10}, 275 (1974).

\bibitem{Georgi:sy}
H.~Georgi and S.~L.~Glashow,
Phys.\ Rev.\ Lett.\  {\bf 32}, 438 (1974).

\bibitem{Arkani-Hamed:2001tb}
N.~Arkani-Hamed, T.~Gregoire and J.~Wacker,
arXiv:hep-th/0101233.

\bibitem{Babu:1993we}
K.~S.~Babu and S.~M.~Barr,
Phys.\ Rev.\ D {\bf 48}, 5354 (1993)
[arXiv:hep-ph/9306242].

\bibitem{Giudice:1988yz} G.~F.~Giudice and A.~Masiero,

\bibitem{leptogenesis}
For a recent review see: W.~Buchmuller,
arXiv:hep-ph/0204288.

\bibitem{Babu:1995hr}
K.~S.~Babu and S.~M.~Barr,
Phys.\ Lett.\ B {\bf 381}, 202 (1996)
[arXiv:hep-ph/9511446].

\bibitem{Sato:1997hv}
J.~Sato and T.~Yanagida,
Phys.\ Lett.\ B {\bf 430}, 127 (1998)
[arXiv:hep-ph/9710516].

\bibitem{Albright:1998vf}
C.~H.~Albright, K.~S.~Babu and S.~M.~Barr,
Phys.\ Rev.\ Lett.\  {\bf 81}, 1167 (1998)
[arXiv:hep-ph/9802314].

\bibitem{Irges:1998ax}
N.~Irges, S.~Lavignac and P.~Ramond,
Phys.\ Rev.\ D {\bf 58}, 035003 (1998)
[arXiv:hep-ph/9802334].

\bibitem{Kaplan:1999ac}
D.~E.~Kaplan, G.~D.~Kribs and M.~Schmaltz,
Phys.\ Rev.\ D {\bf 62}, 035010 (2000)
[arXiv:hep-ph/9911293].

\bibitem{Chacko:1999mi}
Z.~Chacko, M.~A.~Luty, A.~E.~Nelson and E.~Ponton,
JHEP {\bf 0001}, 003 (2000)
[arXiv:hep-ph/9911323].

\bibitem{Arkani-Hamed:1998}
N.~Arkani-Hamed, S.~Dimopoulos,
Phys.\ Rev.\ D {\bf 65}, 052003 (2002)
[arXiv:hep-ph/9811353].

\bibitem{Arkani-Hamed:1999pv}
N.~Arkani-Hamed, L.~J.~Hall, D.~R.~Smith and N.~Weiner,
Phys.\ Rev.\ D {\bf 63}, 056003 (2001)
[arXiv:hep-ph/9911421].

\bibitem{Baer:2002by}
H.~Baer, C.~Balazs, A.~Belyaev, R.~Dermisek, A.~Mafi and A.~Mustafayev,
arXiv:hep-ph/0204108.

\end{thebibliography}
\end{document}